\documentclass [twocolumn,showpacs,preprintnumbers,amsmath,amssymb,prb,floatfix] {revtex4}

\usepackage{epsfig,psfrag}
\usepackage{dcolumn}
\usepackage{bm}
\usepackage{graphicx}
\usepackage{color}

\begin{document}

\title{Low-energy theory and RKKY interaction for interacting
quantum wires with Rashba spin-orbit coupling}
\author{Andreas Schulz$^1$, Alessandro De Martino$^2$,
Philip Ingenhoven$^{1,3}$, and Reinhold Egger$^1$}
\affiliation{$^1$~Institut f\"ur Theoretische Physik,
Heinrich-Heine-Universit\"at, D-40225  D\"usseldorf, Germany\\
$^2$~Institut f\"ur Theoretische Physik,
Universit\"at zu K\"oln, Z\"ulpicher Strasse 77, D-50937 K\"oln,  
Germany\\
$^3$~Institute of Fundamental Sciences,
Massey University, Private Bag 11 222, Palmerston North, New Zealand}
\date{\today}

\begin{abstract}
We present the effective low-energy theory for interacting 1D
quantum wires subject to Rashba spin-orbit coupling.
Under a one-loop renormalization group scheme including all allowed
interaction processes for not too weak Rashba coupling,
we show that electron-electron backscattering is
an irrelevant perturbation.  Therefore no gap arises and
electronic transport is described by a modified
Luttinger liquid theory.  As an application of the theory, we
discuss the RKKY interaction between two magnetic impurities.
Interactions are shown to induce a slower power-law decay
of the RKKY range function
than the usual 1D noninteracting $\cos(2k_F x)/|x|$ law.
Moreover, in the noninteracting Rashba wire, the spin-orbit coupling
causes a twisted (anisotropic) range function with
several different spatial oscillation periods. In the interacting case,
we show that one special oscillation period leads to the slowest  
decay, and
therefore dominates the RKKY interaction for large separation.
\end{abstract}

\maketitle

\section{Introduction}
\label{sec1}
Spin transport in one-dimensional (1D) quantum wires continues
to be a topic of much interest in solid-state and nanoscale physics,
offering interesting fundamental questions as well as
technological applications.\cite{fabian}  Of particular
interest to this field is
the spintronic field effect transistor (spin-FET) proposal
by Datta and Das,\cite{datta} where a gate-tunable Rashba
spin-orbit interaction (SOI) of strength $\alpha$ allows
for a purely electrical manipulation of the spin-dependent current.
While the Rashba SOI arises from a structural
inversion asymmetry\cite{rashba,winkler,winklerbook} of the two- 
dimensional
electron gas  (2DEG) in semiconductor devices hosting the quantum wire,
additional sources for SOI can be present.
In particular, for bulk inversion asymmetric materials, the
Dresselhaus SOI (of strength $\beta$) should also be
taken into account.  By tuning the Rashba SOI (via gate voltages)
to the special point  $\alpha=\beta$, the
spin-FET was predicted to show a remarkable insensitivity
to disorder,\cite{schliemann}
see also Ref.~\onlinecite{bernevig}. On top of these two, additional
(though generally weaker) contributions may arise from the
electric confinement fields  forming the quantum wire.
In this paper, we focus on the case of Rashba
SOI and disregard all other SOI terms.
This limit can be realized experimentally by applying sufficiently
strong backgate voltages,\cite{exp1,exp1b,grundler,exp2}
which create  a large interfacial electric field and
hence a significant and tunable Rashba SOI coupling $\alpha$.
The model studied below may also be relevant to
1D electron surface states of self-assembled gold chains.\cite{schaefer}

The noninteracting theory of such a ``Rashba quantum wire'' has been
discussed in the literature,\cite 
{moroz1,moroz2,governale02,silva,loss2,dario}
and is summarized in Sec.~\ref{sec2} below.
We here discuss electron-electron (e-e) interaction effects
in the 1D limit, where only the lowest (spinful) band is occupied.
The bandstructure at low energy scales
is then characterized by two velocities,\cite{tsvelik}
\begin{equation}\label{vab}
v_{A,B} =  v_F ( 1\pm \delta ), \quad \delta(\alpha) \propto \alpha^4 .
\end{equation}
These reduce to a single Fermi velocity $v_F$ in the absence of Rashba
SOI ($\delta=0$ for $\alpha=0$),
but they will be different for $\alpha\ne 0$, reflecting the broken
spin $SU(2)$ invariance in a spin-orbit coupled system.
The small-$\alpha$ dependence $\delta\propto \alpha^4$
follows for the model below and has also been
reported in Ref.~\onlinecite{haus1}.
Therefore, the velocity splitting (\ref{vab}) is typically weak.
While a similar velocity splitting also happens in a magnetic
Zeeman field (without SOI),\cite{zeeman} the underlying physics is
different since time-reversal symmetry is not broken by SOI.

The bandstructure of a single-channel quantum wire
with Rashba SOI should be obtained by taking into account at least the
lowest two (spinful) subbands, since a
restriction to the lowest subband alone would eliminate spin
relaxation.\cite{governale02,governale04,kaneko}
The problem in this truncated Hilbert space can
be readily diagonalized, and yields two pairs of energy bands.
When describing a single-channel quantum wire, one then keeps only
the lower pair of these energy bands.
We mention in passing that bandstructure effects in the presence of
both Rashba SOI and magnetic fields have also been
studied.\cite{pershin,schapers,pereira,debald,serra}
In addition, the possibility of a spatial modulation of the Rashba  
coupling
was discussed,\cite{wang} but such phenomena will not be further
considered here.  Finally, disorder effects were addressed in
Refs.~\onlinecite{kettemann,scheid}.

For 1D quantum wires, it is well known that the inclusion of e-e
interactions leads to a breakdown of Fermi liquid theory, and often
implies Luttinger liquid (LL) behavior.
This non-Fermi liquid state of matter has a number of
interesting features, including the phenomenon of spin-charge
separation.\cite{gogolin} Motivated mainly by the question of how the
Rashba spin precession and Datta-Das oscillations
in spin-dependent transport are affected by e-e interactions,
Rashba SOI effects on electronic transport in interacting quantum wires
have been studied in recent
papers.\cite 
{governale02,governale04,moroz3,moroz4,epl,haus1,iucci,haus2}
In effect, however, all those works only took e-e forward scattering  
processes
into account. Because of the Rashba SOI, one obtains
a modified LL phase with broken spin-charge separation,\cite 
{moroz3,moroz4}
leading to a drastic influence on observables such as the
spectral function or the tunneling density
of states.  Moroz \textit{et al.}~argued that e-e backscattering  
processes are
irrelevant in the renormalization group (RG) sense, and hence
can be omitted in a low-energy theory.\cite{moroz3,moroz4}
Unfortunately, their theory relies on
an incorrect spin assignment of the subbands,\cite 
{governale02,governale04}
which then invalidates several aspects of their
treatment of interaction processes.

The possibility that e-e backscattering processes become relevant (in  
the
RG sense) in a Rashba quantum wire was
raised in Ref.~\onlinecite{gritsev}, where a spin gap
was found under a weak-coupling two-loop RG scheme. If valid,
this result has important consequences for the physics of such systems,
and would drive them into a spin-density-wave type state.
To establish the spin gap, Ref.~\onlinecite{gritsev}
starts from a strict 1D single-band model and assumes both
  $\alpha$ and the e-e interaction as weak
coupling constants flowing under the RG.  Our approach below
is different in that we include the Rashba coupling $\alpha$
from the outset in the single-particle sector, i.e., in a  
nonperturbative
manner. We then consider the one-loop RG flow of all possible
interaction couplings allowed by momentum conservation (for not
too small $\alpha$).
This is an important difference to the scheme of Ref.~\onlinecite 
{gritsev},
since the Rashba SOI eliminates certain interaction processes
which become momentum-nonconserving. This mechanism is captured by our
approach.  The one-loop RG flow then turns out to be
equivalent to a Kosterlitz-Thouless flow, and for the
initial values realized in this problem, e-e backscattering processes
are always irrelevant.  Our conclusion is therefore that no spin gap
arises because of SOI, and a modified LL picture is always sufficient.
We mention in passing that in the presence of a magnetic field
(which we do not consider), a spin gap can be present because
of spin-nonconserving e-e ``Cooper'' scattering processes;\cite 
{oleg1,oleg2}
the effects of e-e forward scattering in Rashba wires with
magnetic field were studied as well.\cite{yu,lee,cheng,thierry}
Below, we also provide estimates for the \textit{renormalized}
couplings entering the modified LL theory, see Eq.~(\ref{gfix}) below.
When taking bare (instead of renormalized) couplings, we recover
previous results.\cite{governale04} Note that the SOI in
carbon nanotubes\cite{prlale} or graphene ribbons\cite{sandler}
leads to a similar yet different LL description.
In particular, for (achiral) carbon nanotubes,
the leading SOI does not break spin-charge separation.\cite{prlale}
We here only discuss Rashba SOI effects in semiconductor quantum wires
in the absence of magnetic fields.

We apply our formalism to a study of the
Ruderman-Kittel-Kasuya-Yosida (RKKY) interaction\cite{RKKY,RKKYkittel}
between two spin-$1/2$ magnetic impurities, ${\bm \Sigma}_{1,2}$,
separated by a distance $x$.  The RKKY interaction is mediated by the
conduction electrons in the quantum wire which are exchange-coupled
(with  coupling $J$) to the impurity spins.
In the absence of both the e-e interaction and the
SOI, one finds an isotropic exchange (Heisenberg) Hamiltonian,\cite 
{RKKYkittel}
\begin{equation}\label{rkkybasic}
H_{\rm RKKY} = - J^2 F_{\rm ex}(x) \ {\bm \Sigma}_1\cdot {\bm \Sigma}_2,
\quad
F_{\rm ex}(x)\propto \frac{ \cos(2k_F x)}{|x|},
\end{equation}
where the $2k_F$-oscillatory RKKY range function $F_{\rm ex}(x)$
is specified for the 1D case.
When the spin $SU(2)$ symmetry is broken by the SOI, spin precession  
sets in
and the RKKY interaction is generally of a more complicated (twisted)  
form.
For a noninteracting Rashba quantum wire, it has indeed
been established\cite{bruno,lyu,simonin} that the RKKY interaction  
becomes
anisotropic and thus has a tensorial character.
It can always be decomposed into an exchange (scalar)
part, a Dzyaloshinsky-Moriya-like (vector) interaction, and an Ising- 
like
(traceless symmetric tensor) coupling.
On the other hand, in the presence of e-e interactions but without SOI,
the range function has been shown\cite{RKKYreinh}
  to exhibit a slow power-law decay,
$F_{\rm ex}(x)\propto\cos(2k_F x) |x|^{-\eta}$, with an interaction- 
dependent
exponent $\eta<1$.  The RKKY interaction in interacting
quantum wires with SOI has not been studied before.

For the benefit of the focussed reader, we briefly summarize the
main results of our analyis. The effective low-energy
theory of an interacting Rashba quantum wire is given in Eq.~(\ref 
{tll}),
with the velocities (\ref{veloc}) and the dimensionless interaction
parameters (\ref{llpar}).  Previous theories did not fully account  
for the
e-e backscattering processes, and the conspiracy of these processes
with the broken $SU(2)$ invariance due to spin-orbit effects leads
to $K_s<1$ in Eq.~(\ref{llpar}). This in turn implies novel
effects in the RKKY interaction of an interacting Rashba wire.
In particular, the power-law decay exponent
in an interacting Rashba wire, see  Eq.~(\ref{etaB}),
depends explicitly on both the interaction strength and on the
Rashba coupling.

The structure of the remainder of this paper is as follows.
In Sec.~\ref{sec2}, we discuss the bandstructure.
Interaction processes and the one-loop RG scheme are discussed in
Sec.~\ref{sec3}, while the LL description is
provided in Sec.~\ref{sec4}.
The RKKY interaction mediated by an interacting Rashba quantum wire
is then studied in Sec.~\ref{sec5}.  Finally, we offer some  
conclusions in
Sec.~\ref{sec6}.  Technical details can be found in the Appendix.
   Throughout the paper we use units where $\hbar=1$.

\section{Single-particle description}\label{sec2}

\begin{figure}
\scalebox{0.3}{\includegraphics{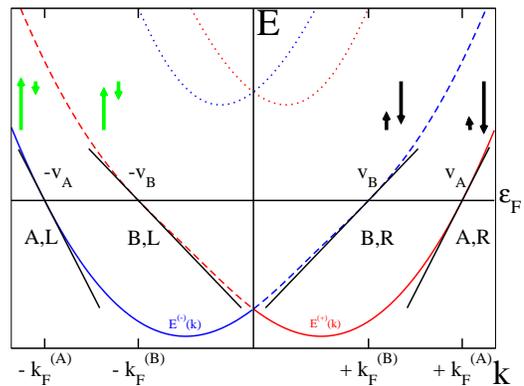}}
\caption{\label{f1} (Color online)
Schematic band structure (\ref{spectrum}) of a typical 1D Rashba  
quantum wire.
The red/blue curves show the $s=\pm$ bands, and the
dotted curves indicate the next subband (the Fermi energy
$\epsilon_F$ is assumed below that band).
For the low-energy description, we linearize the dispersion.
It is notationally convenient to introduce bands A (solid lines)
and B (dashed lines). Green and black arrows indicate the
respective spin amplitudes (exaggerated).
The resulting Fermi momenta are $\pm k_F^{(A,B)}$, with
Fermi velocities $v_{A,B}$.  }
\end{figure}

We consider a quantum wire electrostatically confined in the $z$- 
direction
within the 2DEG ($xz$-plane) by a harmonic potential,
$V_c(z)= m\omega^2 z^2/2$, where $m$ is the effective mass.
The noninteracting problem is then defined by the single-particle
Hamiltonian\cite{moroz1,moroz2,governale02,rashba,loss2}
\begin{equation}\label{h0}
H_{\rm sp} = \frac{1}{2m} \left(p_x^2+p_z^2\right)+ V_c(z) +
\alpha \left( \sigma_z p_x - \sigma_x p_z\right),
\end{equation}
where $\alpha$ is the Rashba coupling and
the Pauli matrices $\sigma_{x,z}$ act in spin space.
For $\alpha=0$, the transverse problem is diagonal in
terms of the familiar 1D harmonic oscillator eigenstates
(Hermite functions) $H_n(z)$, with
$n=0,1,2,\ldots$ labeling the subbands (channels).
Eigenstates of Eq.~(\ref{h0}) have
conserved longitudinal momentum $p_x=k$,
and with the $z$-direction as spin quantization axis, $\sigma_z|\sigma 
\rangle=
\sigma|\sigma\rangle$ with $\sigma=\uparrow,\downarrow=\pm$,
the $\sigma_x p_z$ term implies mixing of adjacent subbands
with associated spin flips. Retaining only the lowest ($n=0$) subband
from the outset thus excludes spin relaxation.
We follow Ref.~\onlinecite{governale02} and keep
the two lowest bands, $n=0$ and $n=1$.
The higher subbands $n\geq 2$ yield only tiny corrections, which
can in principle be included as in Ref.~\onlinecite{loss2}.
The resulting $4\times 4$ matrix representing $H_{\rm sp}$ in
this truncated Hilbert space
is readily diagonalized and yields four energy bands.
We choose the Fermi energy
such that only  the lower two bands, labeled by $s=\pm$, are occupied,
and arrive at a reduced two-band model, where the quantum number
$s=\pm$ replaces the spin quantum number.
The dispersion relation is
\begin{equation} \label{spectrum}
E_{s}(k)=\omega + \frac{k^2}{2m} -
\sqrt{\left(\frac{\omega}{2}+s\alpha k\right)^{2}+
\frac{m\omega \alpha^2}{2}},
\end{equation}
with eigenfunctions $\sim e^{ikx}\phi_{k,s}(z)$.
The resulting asymmetric energy bands (\ref{spectrum}) are
shown in Fig.~\ref{f1}.
The transverse spinors (in spin space) are given by
\begin{eqnarray}\label{eigenfunction}
\phi_{k,+}(z)&=&
\left( \begin{array}{c} i \cos [\theta_+(k)] H_1(z) \\ \sin[\theta_+(k)]
H_0(z) \end{array}\right), \\ \nonumber
\phi_{k,-}(z)&=&
\left( \begin{array}{c}  \sin [\theta_-(k)] H_0(z) \\ i\cos[\theta_-(k)]
H_1(z) \end{array}\right),
\end{eqnarray}
with $k$-dependent spin
rotation angles (we take $0\leq \theta_s(k) \leq \pi/2$)
\begin{equation}\label{theta}
\theta_s(k) = \frac12 \cot^{-1}\left( \frac{- 2 s k- \omega/\alpha}{
  \sqrt{2m\omega} }\right) =\theta_{-s}(-k).
\end{equation}
As a result of subband mixing, the two spinor components of
$\phi_{k,s}(z)$ carry a different $z$-dependence. They
are therefore not just the result of a $SU(2)$ rotation.
For $\alpha=0$, we recover $\theta_s=\pi/2$, corresponding to
the usual spin up and down eigenstates, with $H_0(z)$ as transverse
wavefunction; the $s=+$ ($s=-$) component then describes the
$\sigma=\downarrow$ ($\sigma=\uparrow$) spin eigenstate.
However, for $\alpha\ne 0$,
a peculiar implication of the Rashba SOI follows.
{}From Eq.~(\ref{theta}) we have $\lim_{k\to \pm \infty}
\theta_s(k)=(1\pm s)\pi/4,$
such that both $s=\pm$ states have (approximately) spin $\sigma= 
\downarrow$
for $k\to \infty$ but $\sigma=\uparrow$ for $k\to -\infty$;
the product of spin and chirality thus always
approaches $\sigma {\rm sgn}(k)=-1$.
Moreover, under the time-reversal transformation,
${\cal T}=i\sigma_y {\cal C}$ with the  complex conjugation
operator ${\cal C}$, the two subbands are exchanged,
\begin{equation}\label{trs}
  e^{-ikx} \phi_{-k,-s}(z) = s {\cal T}[e^{ikx} \phi_{k,s}(z)] ,
\quad  E_{-s}(-k)  = E_s(k).
\end{equation}
Time-reversal symmetry, preserved in the truncated description,
makes this two-band model of a Rashba quantum wire
qualitatively different from Zeeman-spin-split models.\cite{zeeman}

In the next step, since we are interested in the low-energy physics,
we linearize the dispersion relation around the Fermi points $\pm k_F^ 
{(A,B)}$,
see Fig.~\ref{f1}, which results in two velocities $v_A$ and $v_B$, see
Eq.~(\ref{vab}).
The linearization of the dispersion relation of multi-band quantum wires
around the Fermi level is known to be an excellent
approximation for weak e-e interactions.\cite{gogolin}
Explicit values for $\delta$ in Eq.~(\ref{vab})
can be derived from Eq.~(\ref{spectrum}), and we find $\delta(\alpha) 
\propto
\alpha^4$ for $\alpha\to 0$,
  in accordance with previous estimates.\cite{haus1}  We mention that
$\delta\alt 0.1$ has been estimated for typical geometries
in Ref.~\onlinecite{moroz4}.
The transverse spinors $\phi_{ks}(z)$, Eq.~(\ref{eigenfunction}),  
entering the
low-energy description can be taken at $k=\pm k_F^{(A,B)}$, where
the spin rotation angle (\ref{theta}) only assumes one of the two values
\begin{equation}\label{angleab}
\theta_A= \theta_+\left ( k_F^{(A)}\right ),\quad \theta_B=\theta_- 
\left (
k_F^{(B)} \right ) .
\end{equation}
The electron field operator $\Psi(x,z)$ for
the linearized two-band model with $\nu=A,B=+,-$ can then be expressed
in terms of 1D fermionic field operators $\psi_{\nu,r}(x)$,
where $r=R,L=+,-$ labels right- and left-movers,
\begin{equation}\label{expand}
\Psi(x,z) = \sum_{\nu,r=\pm} e^{i r k_F^{(\nu)}x } \
\phi_{r k_F^{(\nu)} , s= \nu r }(z) \
   \psi_{\nu,r}(x),
\end{equation}
with $\phi_{k,s}(z)$  specified in
Eq.~(\ref{eigenfunction}).  Note that in the left-moving sector, band  
indices
have been interchanged according to the labeling in Fig.~\ref{f1}.

In this way, the noninteracting
second-quantized Hamiltonian takes the standard form
for two inequivalent species of 1D massless Dirac fermions
with different velocities,
\begin{equation}\label{dirac}
H_0 = -i \sum_{\nu,r=\pm} r v_\nu \int dx\ \psi^\dagger_{\nu,r}  
\partial_x
\psi^{}_{\nu,r}.
\end{equation}
The velocity difference implies the breaking
of the spin $SU(2)$ symmetry, a direct consequence of
SOI. For $\alpha=0$, the index $\nu$ coincides with the spin
quantum number $\sigma$ for left-movers
and with $-\sigma$ for right-movers, and
the above formulation reduces to the usual Hamiltonian
for a spinful single-channel quantum wire.

\section{Interaction effects}\label{sec3}

Let us now include e-e interactions in such a single-channel
disorder-free Rashba quantum wire. With the
expansion (\ref{expand}) and
${\bf r}=(x,z)$, the second-quantized two-body Hamiltonian
\begin{equation}
H_I=\frac12\int d{\bf r}_1 d{\bf r}_2 \ \Psi^\dagger({\bf r}_1)
\Psi^\dagger({\bf r}_2) V({\bf r}_1-{\bf r}_2) \Psi({\bf r}_2) \Psi 
({\bf r}_1)
\end{equation}
leads to 1D interaction processes. We here assume that the
e-e interaction potential $V({\bf r}_1-{\bf r}_2)$
  is externally screened, allowing to
describe the 1D interactions as effectively local.  Following
standard arguments, for weak e-e interactions,
going beyond this approximation at most leads to irrelevant
corrections.\cite{footnote}  We then obtain the local 1D interaction
Hamiltonian\cite{starykh}
\begin{equation} \label{1dint}
H_I =\frac{1}{2} \sum_{ \{ \nu_i, r_i \} } V_{ \{ \nu_i, r_i\} }
\int dx \ \psi^\dagger_{\nu_1,r_1} \psi^\dagger_{\nu_2,r_2}
\psi^{}_{\nu_3,r_3} \psi^{}_{\nu_4,r_4} ,
\end{equation}
where the summation runs over all quantum numbers $\nu_1,\ldots,\nu_4$
and $r_1,\ldots,r_4$ subject to momentum conservation,
\begin{equation}\label{consmom}
r_1 k_F^{(\nu_1)} +r_2 k_F^{(\nu_2)}= r_3 k_F^{(\nu_3)} +r_4 k_F^ 
{(\nu_4)} .
\end{equation}
With the momentum transfer $q= r_1 k_F^{(\nu_1)}-r_4 k_F^{(\nu_4)}$ and
the partial Fourier transform
\begin{equation}\label{pot}
\tilde V(q;z)=\int dx \ e^{-iqx} V(x,z)
\end{equation}
of the interaction potential,
the interaction matrix elements in Eq.~(\ref{1dint}) are given by
\begin{eqnarray}
V_{\{ \nu_i,r_i\}} &=& \int dz_1 dz_2 \ \tilde V(q;z_1-z_2)   
\nonumber \\
  &\times& \ \left[ \phi^\dagger_{r_1 k_F^{(\nu_1)} , \nu_1 r_1 }  \cdot
\phi_{r_4 k_F^{(\nu_4)} , \nu_4 r_4 }\right](z_1) \nonumber \\ &\times&
  \left [ \phi^\dagger_{r_2 k_F^{(\nu_2)} , \nu_2 r_2 }  \cdot
\phi_{r_3 k_F^{(\nu_3)} , \nu_3 r_3 }\right](z_2). \label{intmat}
\end{eqnarray}
Since the Rashba SOI produces a splitting of the Fermi momenta for the
two bands, $\left| k^{(A)}_F-k_F^{(B)}\right| \simeq 2\alpha m$,
the condition (\ref{consmom}) eliminates one important interaction  
process
available for $\alpha=0$, namely interband backscattering (see below).
This is a distinct SOI effect besides the broken spin $SU(2)$
invariance.  Obtaining the complete ``g-ology'' classification\cite 
{gogolin}
of all possible interaction processes
allowed for $\alpha\ne 0$ is then a straightforward exercise.
The corresponding values of the interaction matrix elements are
generally difficult to evaluate explicitly, but in the most important
case of a thin wire,
\begin{equation}\label{thinwire}
  d \gg \frac{1}{\sqrt{m\omega}},
\end{equation}
where $d$ is the screening length (representing, e.g.,
the distance to a backgate), analytical expressions
can be obtained.\cite{footnoteDelta}
To simplify the analysis and allow for
analytical progress, we therefore employ the thin-wire
approximation (\ref{thinwire}) in what follows.
In that case, we can neglect the $z$ dependence in Eq.~(\ref{pot}).
Going beyond this approximation would only imply slightly modified
values for the e-e interaction couplings used below.
Using the identity
\begin{eqnarray}\label{ident}
&& \int dz \ \left[ \phi^\dagger_{r k_F^{(\nu)} , \nu r }  \cdot
\phi_{r' k_F^{(\nu')} , \nu' r' }\right](z) = \\ \nonumber
&& =
\delta_{\nu\nu'}\delta_{rr'}+\cos(\theta_A-\theta_B)\delta_{\nu,-\nu'}
\delta_{r,-r'},
\end{eqnarray}
where the angles $\theta_{A,B}$ were specified in Eq.~(\ref{angleab}),
only two different values $W_0$ and $W_1$ for the matrix elements
  in Eq.~(\ref{intmat}) emerge.  These nonzero matrix elements are
\begin{eqnarray}
  \nonumber
&& V_{ \nu r, \nu' r', \nu' r', \nu r} \equiv W_0 = \tilde V(q=0) , \\
\label{g01} && V_{ \nu r, \nu' r', -\nu' -r', -\nu -r} \equiv W_1 \\
\nonumber && = \cos^2(\theta_A-\theta_B) \ \tilde V\left(
q=k_F^{(A)}+k_F^{(B)}\right)  .
\end{eqnarray}

We then introduce 1D chiral fermion densities
$\rho_{\nu r}(x)= \ :\psi^\dagger_{\nu r} \psi^{}_{\nu r}:$,
where the colons indicate normal-ordering.
The interacting 1D Hamiltonian is $H=H_0+H_I$
with Eq.~(\ref{dirac}) and
\begin{eqnarray} \nonumber
H_I &=& \frac{1}{2} \sum_{\nu \nu', r r'} \int dx \Bigl(
  [ g_{2\parallel\nu} \delta_{\nu,\nu'} + g_{2\perp}
\delta_{\nu,-\nu'}]\delta_{r,-r'} \\ & +& \label{1dint1}
[ g_{4\parallel\nu} \delta_{\nu,\nu'} +
g_{4\perp} \delta_{\nu,-\nu'} ]
  \delta_{r,r'} \Bigr)  \ \rho_{\nu r} \rho_{\nu' r'}
\\ \nonumber &+& \frac{g_f}{2} \sum_{\nu r} \int dx \
\psi_{\nu r}^\dagger \psi_{\nu,- r}^\dagger
\psi_{-\nu r}^{}  \psi_{-\nu, -r}^{}.
\end{eqnarray}
The e-e interaction couplings are denoted in analogy to the
standard $g$-ology, whereby the $g_4$
($g_2$) processes describe forward scattering
of 1D fermions with equal (opposite) chirality $r=R,L=+,-$,
and the labels $\parallel$, $\perp$, and $f$ denote
intraband, interband, and band flip processes, respectively.
Since the bands $\nu=A,B=+,-$ are inequivalent, we keep track of the
band index in the intraband couplings.
The $g_f$ term corresponds to intraband backscattering with
band flip. The interband backscattering without band flip is
strongly suppressed since it does not conserve total
momentum\cite{footnote2} and is neglected in the following.
For $\alpha=0$, the $g_{4,\parallel/\perp}$ couplings
coincide with the usual ones\cite{gogolin} for spinful electrons, while
$g_f$ reduces to $g_{1\perp}$ and
$g_{2,\parallel/\perp}\to g_{2,\perp/\parallel}$
due to our exchange of band indices in the left-moving sector.
According to Eq.~(\ref{g01}), the bare values of these
coupling constants are
\begin{eqnarray}
&& g_{4\parallel\nu} = g_{4\perp}=g_{2\parallel\nu}= W_0, \nonumber \\
&& g_{2\perp} =W_0-W_1,\quad g_{f}=W_1.
\label{init}
\end{eqnarray}
The equality of the intraband coupling constants for the two bands
is a consequence of the thin-wire approximation, which also
eliminates certain exchange matrix elements.

The Hamiltonian $H_0+H_I$ then corresponds to a specific realization
of a general asymmetric two band-model, where the one-loop RG  
equations are
known.\cite{starykh,muttalib}  Using RG invariants, we arrive
after some algebra at the two-dimensional Kosterlitz-Thouless
RG flow equations,
\begin{equation}\label{ktrg}
\frac{d\bar g_2}{dl} =  -\bar g_{f}^2,\quad
\frac{d\bar g_{f}}{dl} = - \bar g_{f}  \bar g_2,
\end{equation}
for the rescaled couplings
\begin{eqnarray}\label{scaling}
\bar g_2 &=& \frac{g_{2\parallel A}}{2\pi v_A} +
\frac{g_{2\parallel B}}{2\pi v_B} - \frac{g_{2\perp}}{\pi v_F} ,\\  
\nonumber
\bar g_{f}  &=&  \sqrt{\frac{1+\gamma}{2}}\frac{g_{f}}{\pi v_F} ,
\end{eqnarray}
where we use the dimensionless constant
\begin{equation}\label{gamma}
\gamma=\frac{v_F^2}{v_A v_B}=\frac{1}{1-\delta^2} \geq 1.
\end{equation}
As usual, the $g_4$ couplings do not contribute to the one-loop RG  
equations.
The initial values of the couplings can be read off from Eq.~(\ref 
{init}),
\begin{eqnarray}
\bar g_2(l=0) &=& \frac{(\gamma-1) W_0+W_1}{\pi v_F}, \nonumber \\
  \bar g_{f}(l=0) &=& \sqrt{\frac{1+\gamma}{2}}
  \frac{W_1}{\pi v_F} .\label{init2}
\end{eqnarray}
The solution of Eq.~(\ref{ktrg}) is textbook material,\cite{gogolin} and
$\bar g_f$ is known to be marginally irrelevant
for all initial conditions with
$|\bar g_{f}(0)|\leq \bar g_2(0)$. Using Eqs.~(\ref{g01})
and (\ref{init2}), this implies
with $\gamma\simeq 1+\delta^2$ the condition
\begin{equation}\label{finalinit}
\tilde V(0) \ge \frac{1}{4} \cos^2(\theta_A-\theta_B) \ \tilde V
\left(k_F^{(A)}+k_F^{(B)}\right),
\end{equation}
which is satisfied for all physically relevant repulsive e-e
  interaction potentials.  As a consequence,
intraband backscattering processes with band flip,
described by the coupling $\bar g_{f}$, are \textit{always marginally
irrelevant}, i.e., they flow to zero coupling as the energy scale is  
reduced,
$\bar g_{f}^*= \bar g_{f}(l\to \infty)=0$.
Therefore \textit{no gap arises}, and a modified LL model is the
appropriate low-energy theory.  We mention in passing that even if we
neglect the velocity difference in Eq.~(\ref{vab}), no spin gap is  
expected
in a Rashba wire, i.e., the broken $SU(2)$
invariance in our model is not required to establish the absence of a  
gap.

The above RG procedure also allows us to extract \textit{renormalized
couplings} entering the low-energy LL description.
The fixed-point value $\bar g^*_2=\bar g_2(l\to \infty)$ now
depends on the Rashba SOI through $\gamma$ in Eq.~(\ref{gamma}).
With the interaction matrix elements $W_{0,1}$ in Eq.~(\ref{g01}),
it is given by
\begin{equation} \label{gfix}
\bar g_2^* = \frac{ \sqrt{ [(\gamma-1)W_0+W_1]^2- (\gamma+1)W_1^2/2 }}
{\pi v_F}.
\end{equation}
For $\alpha=0$, we have $\gamma=1$ and therefore $\bar g_2^*=0$.
The Rashba SOI produces the nonzero fixed-point value (\ref{gfix}),
reflecting the broken $SU(2)$ symmetry.

\section{Luttinger liquid description}\label{sec4}

In this section, we describe the resulting effective
low-energy Luttinger liquid (LL) theory of an
interacting single-channel Rashba wire.
Employing Abelian bosonization,\cite{gogolin} we introduce a
boson field and its conjugate momentum for each band $\nu=A,B=+,-$.
It is useful to switch to symmetric
(``charge''), $\Phi_c(x)$ and $\Pi_c(x)=-\partial_x\Theta_c(x)$,
and antisymmetric (``spin'' for $\alpha=0$), $\Phi_s(x)$ and
$\Pi_s(x)=-\partial_x\Theta_s$, linear combinations of these
fields and their momenta.
The dual fields $\Phi$ and $\Theta$ then allow to express the electron
operator from Eq.~(\ref{expand}) and the
``bosonization dictionary,''
\begin{eqnarray}\label{bosonize}
\Psi(x,z) &=& \sum_{\nu,r} \phi_{r k_F^{(\nu)} , \nu r }(z)
\frac{\eta_{\nu r}}{\sqrt{2\pi a}}  \\ &\times& \nonumber
e^{i r k_F^{(\nu)}x +i\sqrt{\pi/2 } [r\Phi_c+\Theta_c+\nu r\Phi_s+
\nu \Theta_s] } ,
\end{eqnarray}
where $a$ is a small cutoff length and $\eta_{\nu r}$ are the standard
  Klein factors.\cite{gogolin,RKKYreinh,conbos}
(To recover the conventional expression for $\alpha=0$, due to our
convention for the band indices in the left-moving sector,
one should replace $\Phi_s,\Theta_s \rightarrow -\Theta_s, - \Phi_s$.)
Using the identity (\ref{ident}), we can now express the 1D charge
and spin densities,
\begin{equation}\label{1ddensities}
\rho(x)=\int dz \, \Psi^\dagger \Psi, \quad
{\bm S}(x)=\int dz \, \Psi^\dagger \frac{\bm \sigma}{2} \Psi,
\end{equation}
in bosonized form. The (somewhat lengthy) result can be found in
Appendix \ref{appa}.

The low-energy Hamiltonian is then taken with the fixed-point
values for the interaction constants, i.e.,
backscattering processes are disregarded and only appear via
the renormalized value of $\bar g_2^*$ in Eq.~(\ref{gfix}).
Following standard steps, the kinetic term $H_0$ and the
forward scattering processes then lead to
the exactly solvable Gaussian field theory of a modified (extended)
Luttinger liquid,
\begin{eqnarray}\label{tll}
H&=& \sum_{j=c,s} \frac{v_j}{2} \int dx \ \left(
K_j \Pi_j^2 + \frac{1}{K_j}(\partial_x \Phi_j)^2 \right) \\
\nonumber &+& v_\lambda \int dx \ \left(
  K_\lambda \Pi_c \Pi_s + \frac{1}{K_\lambda}
(\partial_x \Phi_c) (\partial_x\Phi_s) \right).
\end{eqnarray}
Using the notations $\bar g_4=W_0/\pi v_F$ and
\begin{eqnarray*}
y_\delta &=& \frac{g^*_{2\parallel A} - g^*_{2\parallel B}}{4\pi v_F},\\
y_{\pm} &=& \frac{ g^*_{2\parallel A} +
g^*_{2\parallel B} \pm 2 g^*_{2\perp} }{4\pi v_F },
\end{eqnarray*}
where explicit (but lengthy) expressions for the fixed-point
values $g^*_{2\parallel A/B}$ and $g^*_{2\perp}$
can be straightforwardly obtained from Eqs.~(\ref{scaling}) and (\ref 
{gfix}),
the renormalized velocities appearing in Eq.~(\ref{tll}) are
\begin{eqnarray} \nonumber
v_c &=& v_F \sqrt{(1+ \bar g_{4})^2  - y_{+}^{2}} \\ \nonumber
&\simeq& v_F \sqrt{\left(1+ \frac{W_0}{\pi v_F}\right)^2-\left(
\frac{2W_0-W_1}{2\pi v_F}\right)^2 },  \\ \label{veloc}
v_s &=& v_F \sqrt{1  - y_{-}^{2} } \simeq v_F,\\ \nonumber
v_\lambda &=& v_F  \sqrt{\delta^2 - y_\delta^2} \simeq v_F\delta
\sqrt{1-\left( \frac{W_1}{4\pi v_F} \right)^2 } .
\end{eqnarray}
In the respective second equalities, we have specified the leading
terms in $|\delta|\ll 1$, since the  SOI-induced  relative
velocity asymmetry $\delta$ is small even for rather large $\alpha$,
see  Eq.~(\ref{vab}).
The corrections to the quoted expressions are of ${\cal O}(\delta^2)$
and are negligible in practice.  It is noteworthy that the
``spin'' velocity $v_s$ is \textit{not}\ renormalized for a Rashba wire,
although it is well-known that $v_s$ will be renormalized due to $W_1$
for $\alpha=0$.\cite{gogolin}
This difference can be traced to our thin-wire approximation (\ref 
{thinwire}).
When releasing this approximation, there will be a renormalization in  
general.
Finally, the dimensionless LL interaction parameters in
Eq.~(\ref{tll}) are given by
\begin{eqnarray}\nonumber
K_c &=&\sqrt{\frac{1+ \bar g_{4} - y_{+} }{1+ \bar g_{4}  + y_{+} }}
\simeq \sqrt{\frac{2\pi v_F+W_1}{2\pi v_F+4W_0-W_1}} , \\
\label{llpar}
K_s &=&\sqrt{\frac{1 - y_{-} }{1+ y_{-} }} \simeq
1- \frac{\sqrt{W_0 W_1}}{\sqrt{2}\ \pi v_F} |\delta| , \\ \nonumber
K_\lambda &=& \sqrt{\frac{\delta - y_\delta}{\delta + y_\delta}} \simeq
\sqrt{\frac{4\pi v_F+W_1}{4\pi v_F-W_1}},
\end{eqnarray}
where the second equalities again hold up to contributions
  of ${\cal O}(\delta^2)$.  When the $2k_F$ component
of the interaction potential $W_1=0$, see Eq.~(\ref{g01}),
we obtain $K_s=K_\lambda=1$, and thus recover the theory
of Ref.~\onlinecite{governale04}.  The broken spin $SU(2)$ symmetry
is reflected in $K_s<1$ when both $\delta\ne 0$ and $W_1\ne 0$.

Since we arrived at a Gaussian field theory, Eq.~(\ref{tll}),
all low-energy correlation functions can now be computed analytically  
without
further approximation. The linear algebra problem needed for this
diagonalization is discussed in App.~\ref{appa}.

\section{RKKY interaction}\label{sec5}

Following our discussion in Sec.~\ref{sec1},
we now investigate the combined effects of the Rashba SOI
and the e-e interaction on the RKKY range function.
We include the exchange coupling,
$H'=J \sum_{i=1,2}  {\bm \Sigma}_i \cdot {\bm S}(x_i)$,
of the 1D conduction electron spin density ${\bm S}(x)$ to
localized spin-1/2 magnetic impurities, separated by $x=x_1-x_2$.
The RKKY interaction $H_{\rm RKKY}$, describing spin-spin interactions
  between the two magnetic impurities, is then
obtained by perturbation theory in $J$.\cite{RKKYkittel}
In the simplest 1D case (no SOI, no interactions), it is given by
Eq.~(\ref{rkkybasic}).  In the general case, one can always express
it in the form
\begin{equation}\label{RkkyHam}
H_{\rm RKKY} = - J^2 \sum_{a,b} F^{ab}(x) \Sigma^a_1 \Sigma^b_2,
\end{equation}
with the range function now appearing as a tensor
($\beta=1/k_B T$ for temperature $T$),
\begin{equation}\label{range}
F^{ab}(x) = \int_0^\beta d\tau \ \chi^{ab}(x,\tau).
\end{equation}
Here, the imaginary-time ($\tau$) spin-spin correlation function  
appears,
\begin{equation}\label{spinspin}
\chi^{ab}(x,\tau) = \langle S^a(x,\tau) S^b(0,0) \rangle .
\end{equation}
The 1D spin densities $S^a(x)$ (with $a=x,y,z$)
were defined in Eq.~(\ref{1ddensities}),
and their bosonized expression is given in App.~\ref{appa},
which then allows to compute the correlation functions (\ref{spinspin})
using the unperturbed ($J=0$) LL model (\ref{tll}).
The range function thus effectively coincides with the
static space-dependent spin susceptibility tensor.
When spin $SU(2)$ symmetry is realized, $\chi^{ab}(x)=\delta^{ab}F_ 
{\rm ex}(x)$,
and one recovers Eq.~(\ref{rkkybasic}), but in general
this tensor is not diagonal.
For a LL without Rashba SOI, $F_{\rm ex}(x)$ is as
in Eq.~(\ref{rkkybasic}) but with a slow
power-law decay.\cite{RKKYreinh}

If spin $SU(2)$ symmetry is broken, general arguments
imply that Eq.~(\ref{RkkyHam}) can be decomposed into three terms,  
namely
(i) an isotropic exchange scalar coupling, (ii) a
Dzyaloshinsky-Moriya (DM) vector term, and (iii)
an Ising-like interaction,
\begin{eqnarray} \nonumber
H_{\rm RKKY}/J^2 &=& - F_{\rm ex}(x) {\bm \Sigma}_1 \cdot {\bm \Sigma} 
_2 -
{\bm F}_{\rm DM}(x) \cdot \left( {\bm \Sigma}_1 \times {\bm \Sigma}_2  
\right)\\
\label{RKKYgen}
&-& \sum_{a,b} F^{ab}_{\rm Ising}(x) \Sigma^a_1 \Sigma^b_2 ,
\end{eqnarray}
where $F_{\rm ex}(x)  =  \frac{1}{3}\sum_a F^{aa}(x)$.
The DM vector has the components
\[
F_{\rm DM}^c(x) = \frac{1}{2}\sum_{a,b} \epsilon^{cab} F^{ab}(x),
\]
and the Ising-like  tensor
\[
F_{\rm Ising}^{ab}(x)= \frac{1}{2}\left(
F^{ab} + F^{ba}  -\frac{2}{3} \sum_c F^{cc}
\delta^{ab} \right)(x)
\]
is symmetric and traceless.
For a 1D noninteracting quantum wire with Rashba SOI, the ``twisted''  
RKKY
Hamiltonian (\ref{RKKYgen})
has recently been discussed,\cite{bruno,lyu,simonin}
and all range functions appearing in Eq.~(\ref{RKKYgen}) were
shown to decay $\propto |x|^{-1}$, as expected for a noninteracting  
system.
Moreover, it has been emphasized\cite{lyu}
that there are different spatial oscillation periods, reflecting the
presence of different Fermi momenta $k_F^{(A,B)}$
in a Rashba quantum wire.

Let us then consider the extended LL model (\ref{tll}), which includes
the effects of both the e-e interaction and
the Rashba SOI.  The correlation functions (\ref{spinspin})
obey $\chi^{ba}(x,\tau)=\chi^{ab}(-x,-\tau)$, and since we find
$\chi^{xz}=\chi^{yz}=0$, the anisotropy acts only in the $xy$-plane.
The four nonzero correlators are specified in App.~\ref{appa},
where only the long-ranged $2k_F$ oscillatory terms are kept.
These are the relevant correlations determining the RKKY
interaction in the interacting quantum wire.  We
note that in the noninteracting case, there is also a ``slow''
oscillatory component, corresponding to a contribution
to the RKKY range function $\propto
\cos\left[ \left(k_F^{(A)}-k_F^{(B)}\right)x\right]/|x|$.
Remarkably, we find that this $1/x$ decay law is not changed by  
interactions.
However, we will show below that interactions cause a slower decay of  
certain
  ``fast'' oscillatory terms, e.g., the contribution
$\propto \cos(2k_F^{(B)}x)$.
We therefore do not further discuss the ``slow'' oscillatory
terms in what follows.

Collecting everything, we find the various range functions in
Eq.~(\ref{RKKYgen}) for the interacting case,
\begin{eqnarray}  \nonumber
F_{\rm ex}(x) &=&\frac{1}{6}\sum_\nu \Bigl[
\left(1+\cos^2(2\theta_\nu) \right) \cos\left (2k_F^{(\nu)} x\right)
F_\nu^{(1)}(x) \\
\nonumber
&+& \cos^2 (\theta_A+\theta_B)
\cos\left[(k^{(A)}_F+k_F^{(B)})x
\right] F_\nu^{(2)}(x) \Bigr], \\ \label{rangell}
{\bm F}_{\rm DM}(x) &=&  \hat e_z \sum_\nu \frac{\nu}{2}
  \cos (2\theta_\nu) \sin \left(2k_F^{(\nu)}x\right)  F^{(1)}_\nu(x),\\ 
  \nonumber
F^{ab}_{\rm Ising}(x) &=& \left[ \frac12 \sum_\nu G_\nu^a(x) - F_{ex}(x)
\right] \delta^{ab} ,
\end{eqnarray}
with the auxiliary vector
\[
{\bm G}_\nu = \left( \begin{array}{c}
   \cos\left(2k_F^{(\nu)} x\right)   F_\nu^{(1)}(x) \\
   \cos^2(2\theta_\nu) \cos\left(2k_F^{(\nu)} x\right)   F_\nu^{(1)} 
(x) \\
  \cos^2(\theta_A+\theta_B)
\cos\left[(k^{(A)}_F+k_F^{(B)})x
\right] F_\nu^{(2)}(x) \end{array}\right).
\]
The functions $F_\nu^{(1,2)}(x)$ follow by integration over $\tau$
from $\tilde F_\nu^{(1,2)}(x,\tau)$, see  Eqs.~(\ref{func1}) and (\ref 
{func2})
in App.~\ref{appa}.
This implies the respective decay laws for  $a\ll |x|\ll v_F/k_B T$,
\begin{eqnarray}\label{asymp}
F_\nu^{(1)}(x) &\propto& |a/x|^{-1+K_c+K_s+2\nu (1-K_c/K^2_\lambda)
\frac{v_\lambda K_\lambda}{v_c+v_s} },\\
F_\nu^{(2)}(x) &\propto& |a/x|^{-1+K_c+1/K_s}. \nonumber
\end{eqnarray}
All those exponents approach unity in the noninteracting limit,
in accordance with previous results.\cite{bruno,lyu}
Moreover, in the absence of SOI ($\alpha=\delta=0$),
Eq.~(\ref{asymp}) reproduces the known $|x|^{-K_c}$
decay law for the RKKY interaction in a conventional LL.\cite{RKKYreinh}

Since $K_s<1$ for an interacting Rashba wire with $\delta\ne 0$,
see Eq.~(\ref{llpar}), we conclude that $F_\nu^{(1)}$ with $\nu=B$,
corresponding to the slower velocity $v_B=v_F(1-\delta),$
leads to the slowest decay of the RKKY interaction.
For large distance $x$, the RKKY interaction is therefore dominated
by the $2k_F^{(B)}$ oscillatory part, and all range functions
decay $\propto |x|^{-\eta_B}$ with the exponent
\begin{equation}\label{etaB}
\eta_B = K_c + K_s - 1 - 2 \left(1-\frac{K_c}{K_\lambda^2}\right)
\frac{v_\lambda K_\lambda}{v_c+v_s} < 1.
\end{equation}
This exponent depends both on the e-e interaction potential
and on the Rashba coupling $\alpha$. The latter dependence also
implies that electric fields are able to change the power-law
decay of the RKKY interaction in a Rashba wire.
The DM vector coupling also illustrates that the SOI is able
to effectively induce off-diagonal couplings in spin space,
reminiscent of spin precession effects. Also these RKKY couplings
are $2k_F^{(B)}$ oscillatory and show a power-law decay with the
exponent (\ref{etaB}).

\section{Discussion}\label{sec6}

In this paper, we have presented a careful derivation of the low-energy
Hamiltonian of a homogeneous 1D quantum
wire with not too weak Rashba spin-orbit interactions.  We have  
studied the
simplest case (no magnetic field, no disorder, single-channel limit),
and in particular analyzed the possibility for a spin gap to occur
because of electron-electron backscattering processes.
The initial values for the coupling
constants entering the one-loop RG equations were determined, and
for rather general conditions, they are such that
  backscattering is marginally irrelevant and no spin gap opens.
The resulting low-energy theory is a modified Luttinger liquid,
Eq.~(\ref{tll}), which is a Gaussian field theory
formulated in terms of the boson fields $\Phi_c(x)$ and $\Phi_s(x)$ (and
their dual fields).  In this
state, spin-charge separation is violated due to the Rashba coupling,
but the theory still admits exact results for
essentially all low-energy correlation functions.

Based on our  bosonized expressions for the 1D charge and
spin density, the frequency dependence
of various susceptibilities of interest,
e.g., charge- or spin-density wave  correlations,
can then be computed.
As the calculation closely mirrors the one in Refs.~\onlinecite 
{moroz4,epl},
we do not repeat it here.  One can then infer a ``phase diagram'' from
the study of the dominant susceptibilities.  According to our  
calculations,
due to a conspiracy of the Rashba SOI and the e-e interaction,
spin-density-wave correlations in the $xy$ plane are always dominant
for repulsive interactions.

We have studied the RKKY interaction between two magnetic impurities  
in such an
interacting 1D Rashba quantum wire.
On general grounds, the RKKY interaction
  can be decomposed into an exchange term, a DM vector
term, and a traceless symmetric tensor interaction.
For a noninteracting wire, the corresponding
  three range functions have several spatial oscillation periods with
a common overall decay $\propto |x|^{-1}$. We have shown that
interactions modify this picture.
The dominant contribution (characterized by the slowest
power-law decay) to the RKKY range function
is now $2k_F^{(B)}$ oscillatory for all three terms, with the same
exponent $\eta_B<1$, see Eq.~(\ref{etaB}).  This exponent
depends both on the interaction strength and on the Rashba coupling.
This raises the intriguing possibility to tune the power-law exponent
$\eta_B$ governing the RKKY interaction by an electric field, since
$\alpha$ is tunable via a backgate voltage. We stress again
that interactions imply that a single spatial oscillation period  
(wavelength
$\pi/k_F^{(B)}$) becomes dominant, in contrast to the noninteracting
situation where several competing wavelengths are expected.

The above formulation also holds promise for future calculations
of spin transport in the presence of both interactions and Rashba  
spin-orbit
couplings, and possibly with disorder.
Under a perturbative treatment of impurity backscattering,
otherwise exact statements are possible even out of equilibrium.
We hope that our work will motivate further studies along this line.

\acknowledgments

We wish to thank W. H\"ausler and U. Z\"ulicke for helpful discussions.
This work was supported by the SFB TR 12 of the DFG, and by the ESF
network INSTANS.

\begin{appendix}
\section{Bosonization for the extended Luttinger liquid}
\label{appa}

In this appendix, we provide some technical details related
to the evaluation of the spin-spin correlation function under
the extended Luttinger theory (\ref{tll}).  The exact calculation
of such correlations is possible within the bosonization framework,
and requires a diagonalization of Eq.~(\ref{tll}).

The 1D charge and spin densities (\ref{1ddensities})
can be written as the sum of slow and fast (oscillatory) contributions.
Using Eq.~(\ref{ident}), the bosonized form for the 1D charge density is
\begin{eqnarray*}
\rho(x) &=& \sqrt{\frac{2}{\pi}} \ \partial_x \Phi_c -\frac{2i}{\pi a}
  \eta_{AR} \eta_{AL} \cos(\theta_A-\theta_B)  \\ &\times&
\sin\left[ \left(k_F^{(A)}+k_F^{(B)}\right) x+\sqrt{2\pi} \Phi_c \right]
\cos(\sqrt{2\pi} \Theta_s).
\end{eqnarray*}
Similarly, using the identity
\begin{eqnarray*}
  && \int dz\, \left[ \phi^\dagger_{r k_F^{(\nu)} , \nu r } \
{\bm \sigma} \, \phi_{r' k_F^{(\nu')} , \nu' r' } \right] (z)
= \\
&&  \delta_{r,r'} \left(\begin{array}{c}
\cos\left(\theta_{A}-\theta_{B}\right) \delta_{\nu,-\nu'} \\
-i\nu r \,\cos\left(\theta_{A}+\theta_{B}\right) \delta_{\nu,-\nu'} \\
\nu r\,\cos\left(2\,\theta_{\nu}\right) \delta_{\nu,\nu'}
\end{array} \right) + \\
&&+  \delta_{r,-r'} \left( \begin{array}{c}
\delta_{\nu,\nu'}\\ -i\nu r \,
\cos\left(2\,\theta_{\nu} \right) \delta_{\nu,\nu'} \\
\nu r \,\cos\left(\theta_{A}+\theta_{B}\right) \delta_{\nu,-\nu'}
\end{array} \right),
\end{eqnarray*}
the 1D spin density vector has the components
\begin{widetext}
\begin{eqnarray*}
S^x(x) &=& -i\frac{\eta_{AR}\eta_{BR}}{\pi a}
\cos\left(\theta_A-\theta_B\right)
\cos \left[ \left(k_F^{(A)}-k_F^{(B)}\right) x +\sqrt{2\pi} \Phi_s  
\right]
\sin (\sqrt{2\pi} \Theta_s)  \\
&-& i\frac{\eta_{AR}\eta_{AL}}{\pi a}
\cos \left[ \left(k_F^{(A)} + k_F^{(B)}\right) x +\sqrt{2\pi} \Phi_c  
\right]
\sin \left[ \left(k_F^{(A)}-k_F^{(B)}\right) x +\sqrt{2\pi} \Phi_s  
\right] ,
\end{eqnarray*}
\begin{eqnarray*}
S^y(x) &=& i
  \frac{\eta_{AR}\eta_{BR}}{\pi a} \cos\left(\theta_A + \theta_B\right)
  \sin \left[ \left(k_F^{(A)}-k_F^{(B)}\right) x +\sqrt{2\pi} \Phi_s  
\right]
  \sin( \sqrt{2\pi} \Theta_s ) \\ &-& i
\sum_{\nu=A,B=+,-} \nu  \frac{\eta_{\nu R} \eta_{\nu L}}{2\pi a}
\cos( 2\theta_\nu ) \cos \left[2 k_F^{(\nu)} x +\sqrt{2\pi}
\left( \Phi_c+\nu \Phi_s\right) \right] ,
\end{eqnarray*}
\begin{eqnarray*}
S^z(x) &=& \frac{1}{\sqrt{8\pi}} \left[
\left(\cos 2\theta_A +\cos 2\theta_B \right)
\partial_x \Theta_s +
\left( \cos 2\theta_A -\cos 2\theta_B \right)
\partial_x \Theta_c \right]  \\ \nonumber
&-&i \frac{\eta_{AR}\eta_{BL}}{\pi a} \cos(\theta_A+\theta_B)
\cos \left[ \left(k_F^{(A)}+k_F^{(B)} \right) x +\sqrt{2\pi} \Phi_c  
\right]
\sin (\sqrt{2\pi}\Phi_s ).
\end{eqnarray*}
\end{widetext}
Note that while $\partial_x \Phi_c$ is
proportional to the (slow part of the) charge density,
the (slow) spin density is determined by both $c$ and $s$ sectors.

Next we specify the nonzero components of the
imaginary-time spin-spin correlation function $\chi^{ab}(x,\tau)$,
see Eq.~(\ref{spinspin}).
Using the above bosonized expressions, some algebra yields
\[
\chi^{xx}(x,\tau)=
\sum_\nu \frac{\cos\left ( 2k_F^{(\nu)}x \right )}{2(2\pi a)^2}
\tilde F^{(1)}_\nu(x,\tau),
\]
\[
\chi^{yy}(x,\tau)=
\sum_\nu\frac{ \cos^2(2\theta_\nu) \cos \left(2k_F^{(\nu)}x\right)}{2 
(2\pi a)^2}
  \tilde F^{(1)}_\nu(x,\tau),
\]
\begin{eqnarray*}
\chi^{zz}(x,\tau)&=&
  \sum_{\nu r} \frac{ \cos^2(\theta_A+\theta_B) }{2 (2\pi a)^2}  \\
&& \times \cos
\left[\left(k^{(A)}_F+k^{(B)}_F\right)x\right] \tilde
F^{(2)}_{\nu}(x,\tau),
\end{eqnarray*}
and
\[
\chi^{xy}(x,\tau)= \sum_\nu
\frac{ \nu \cos( 2\theta_\nu ) \sin \left(2k_F^{(\nu)}x\right)}{2(2 
\pi a)^2}
  \tilde F^{(1)}_\nu(x,\tau).
\]
Here, the functions $\tilde F^{(1,2)}_{\nu=A,B=+,-}(x,\tau)$ are  
given by
\begin{widetext}
\[
\tilde F^{(1)}_\nu(x,\tau) = \prod_{j=1,2} \left |
\frac{\beta u_j}{\pi a} \sin \left(\frac{ \pi (u_j \tau - i x ) }
{ \beta u_j }  \right)
\right |^{ - \left( \Gamma^{(j)}_{\Phi_c\Phi_c} +
\Gamma^{(j)}_{\Phi_s \Phi_s} + 2 \nu \Gamma^{(j)}_{\Phi_c\Phi_s} 
\right ) }
\]
and
\[ \tilde F^{(2)}_{\nu} (x,\tau)  =  \prod_{j=1,2}
\left | \frac{\beta u_j}{\pi a} \sin
\left( \frac{\pi(u_j\tau-ix)}{\beta u_j} \right)
  \right |^{- \left( \Gamma^{(j)}_{\Phi_c\Phi_c} +
  \Gamma^{(j)}_{\Theta_s \Theta_s}\right ) } \left[
\frac{ \sin \left( \frac{\pi (u_j \tau + i x ) }{ \beta u_j} \right) }
{ \sin \left(\frac{\pi (u_j \tau - i  x) }{ \beta u_j} \right)}
\right]^{\nu \Gamma^{(j)}_{\Phi_c\Theta_s}}.
\]
\end{widetext}
The dimensionless numbers $\Gamma^{(j)}$ appearing in the
exponents follow from the straightforward (but lengthy)
diagonalization of the extended
LL Hamiltonian (\ref{tll}), where the $u_{j}$ are the velocities of the
corresponding normal modes.
With the velocities (\ref{veloc}) and the dimensionless
Luttinger parameters (\ref{llpar}),
the result of this linear algebra problem can be written as
follows.
The normal-mode velocities $u_1$ and $u_2$ are
\begin{eqnarray*}
&& 2u_{j=1,2}^2 = v_c^2+v_s^2+2v_\lambda^2
- (-1)^j \Bigl [ (v_c^2-v_s^2)^2 + \\
&& +4v_\lambda^2\left[ v_c v_s \left( \frac{K_\lambda^2}{K_c K_s}
+ \frac{K_c K_s}{K_\lambda^2}\right) + v_c^2+v_s^2 \right] \Bigr]^{1/2},
\end{eqnarray*}
and the exponents $\Gamma^{(j=1,2)}$ appearing in
$\tilde F^{(1,2)}_\nu(x,\tau)$ are given by
\[
\Gamma_{\Phi_c\Phi_c}^{(j)} =
  \frac{(-1)^j K_c v_c}{u_j(u_1^2-u_2^2)} \left(v_s^2-u_j^2-
\frac{K_\lambda^2  v_\lambda^2 v_s}{K_c K_s v_c} \right) ,
\]
\[
\Gamma_{\Phi_s\Phi_s}^{(j)} =
  \frac{(-1)^j K_s v_s}{u_j(u_1^2-u_2^2)} \left(v_c^2-u_j^2-
\frac{K_\lambda^2  v_\lambda^2 v_c}{K_c K_s v_s} \right) ,
\]
\[
\Gamma_{\Phi_c\Phi_s}^{(j)} =
  \frac{(-1)^j K_\lambda v_\lambda}{u_j(u_1^2-u_2^2)} \left(v_ 
\lambda^2-u_j^2-
\frac{K_c K_s v_s v_c}{K_\lambda^2 } \right) ,
\]
\[
\Gamma_{\Theta_s\Theta_s}^{(j)} =
  \frac{(-1)^j v_s}{K_s u_j(u_1^2-u_2^2)} \left(v_c^2-u_j^2-
\frac{K_c K_s  v_\lambda^2 v_c}{K_\lambda^2 v_s} \right) ,
\]
\[
\Gamma_{\Phi_c\Theta_s}^{(j)} =
  \frac{(-1)^j v_\lambda}{u_1^2-u_2^2} \left(
\frac{K_\lambda}{K_s} v_s + \frac{K_c}{K_\lambda}v_c\right).
\]

Since $|\delta|\ll 1$, we now employ the
simplified expressions for the velocities in Eq.~(\ref{veloc}) and
the Luttinger liquid parameters in Eq.~(\ref{llpar}), which
are valid up to ${\cal O}(\delta^2)$ corrections.
In the interacting case, this yields
for the normal-mode velocities simply $u_1=v_c$ and $u_2=v_s$. (In the
noninteracting limit, the above equation instead yields $u_1=v_A$
and $u_2=v_B$, see Eq.~(\ref{vab}).)
Moreover, the exponents $\Gamma^{(j)}$ simplify to
\[
\Gamma^{(1)}_{\Phi_c\Phi_c} = K_c ,\quad
\Gamma^{(2)}_{\Phi_c\Phi_c} = \Gamma^{(1)}_{\Phi_s\Phi_s} =
\Gamma^{(1)}_{\Theta_s\Theta_s} = 0 ,
\]
\[
\Gamma^{(2)}_{\Phi_s\Phi_s} = K_s , \quad
\Gamma^{(2)}_{\Theta_s\Theta_s} = 1/K_s ,
\]
\[
\Gamma^{(1)}_{\Phi_c\Phi_s} = \frac{v_\lambda}{v_c^2-v_s^2}
( K_\lambda v_c + K_c v_s/K_\lambda ),
\]
\[
\Gamma^{(2)}_{\Phi_c\Phi_s} = -\frac{v_\lambda}{v_c^2-v_s^2}
( K_\lambda v_s + K_c v_c/K_\lambda ),
\]
\[
\Gamma^{(1,2)}_{\Phi_c\Theta_s} = \pm \Gamma^{(2)}_{\Phi_c\Phi_s}.
\]
Collecting everything and taking the zero-temperature limit,
the functions $\tilde F^{(1,2)}_{\nu=\pm}(x,\tau)$ take the form
\begin{eqnarray} \nonumber
&& \tilde F^{(1)}_\nu(x,\tau)=
\left|\frac{v_c \tau-ix}{ a}\right|^{ -K_c-2\nu v_\lambda\frac{K_\lambda
v_c+K_c v_s/K_\lambda}{v_c^2-v_s^2} } \\ &\times& \label{func1}
\left|\frac{v_s \tau-ix}{ a}\right|^{-K_s+2\nu v_\lambda
\frac{K_\lambda v_s+K_c v_c/K_\lambda}{v_c^2-v_s^2}   } ,
\end{eqnarray}
and
\begin{eqnarray} \label{func2}
&& \tilde F^{(2)}_\nu(x,\tau)= \left|\frac{v_c \tau-ix}{a}\right|^{-K_c}
\left|\frac{v_s \tau-ix}{ a}\right|^{-1/K_s} \\ \nonumber
&\times& \left( \frac{(v_s \tau-ix)(v_c\tau+ix)}{(v_s\tau+ix)(v_c\tau- 
ix)}
\right)^{-\nu \frac{v_\lambda (K_\lambda v_s+K_cv_c/K_\lambda)}{v_c^2- 
v_s^2} }.
\end{eqnarray}
The known form of the spin-spin correlations in a LL with $\alpha=0$
is recovered by putting $v_\lambda \propto \delta=0$.

\end{appendix}


\begin{thebibliography}{99}
\bibitem{footnote}
Technically, one performs a gradient expansion in the relative
coordinate $x_1-x_2$. In this procedure, it is crucial
to understand the products of fermionic operators
in the sense of operator product expansions.
Corrections to the leading terms are then irrelevant in the RG sense.
\bibitem{footnoteDelta}
Note that Eq.~(\ref{thinwire}) excludes the case of ultra-local contact
interactions.
\bibitem{footnote2}
It is in principle possible that such a process becomes important
if a collective density readjustment between subbands takes place
in the wire. However, this can only happen for almost equivalent
subbands, see Ref.~\onlinecite{starykh} for a
detailed discussion.  Here we assume that the SOI is strong enough
to guarantee that such a readjustment does not occur.
\end{thebibliography}
\end{document}